\documentclass[linenumber,twocolumn]{aastex631}

\usepackage{amsmath,amsthm,amssymb,amsfonts}
\usepackage{graphicx}
\usepackage{float}
\usepackage{subfigure}
\usepackage{tabularx}
\usepackage{booktabs} 

\begin{document}

\title{Exploring the Current Star Formation Rate and Nebula Ratio\\of Star-Formation Galaxies at  z \textless\ 0.4 with FADO}

\author[0009-0003-7681-3702]{Yaosong Yu}
\affiliation{Jodrell Bank Centre for Astrophysics, Department of Physics and Astronomy, The University of Manchester, Oxford Road, Manchester M13 9PL, UK}
\affiliation{South-Western Institute for Astronomy Research, Yunnan University, Kunming 650500, People's Republic of China}
\author[0009-0006-9345-9639]{Qihang Chen}
\affiliation{School of Physics and Astronomy, Beijing Normal University, Beijing 100875, China}
\affiliation{Institute for Frontier in Astronomy and Astrophysics, Beijing Normal University, Beijing, 102206, China}
\author[0000-0003-1188-9573]{Liang Jing}
\affiliation{School of Physics and Astronomy, Beijing Normal University, Beijing 100875, China}
\affiliation{Institute for Frontier in Astronomy and Astrophysics, Beijing Normal University, Beijing, 102206, China}
\author[0000-0003-2606-6019]{Ciro Pappalardo}
\affiliation{Instituto de Astrofísica e Ciências do Espaço, Universidade de Lisboa – OAL, Tapada da Ajuda, 1349-018 Lisboa, Portugal
e-mail: fc49545@alunos.fc.ul.pt}
\affiliation{Departamento de Física, Faculdade de Ciências da Universidade de Lisboa, Edifício C8, Campo Grande, 1749-016 Lisboa, Portugal}
\author[0000-0002-4793-1051]{Henrique Miranda}
\affiliation{Instituto de Astrofísica e Ciências do Espaço, Universidade de Lisboa – OAL, Tapada da Ajuda, 1349-018 Lisboa, Portugal
e-mail: fc49545@alunos.fc.ul.pt}
\affiliation{Departamento de Física, Faculdade de Ciências da Universidade de Lisboa, Edifício C8, Campo Grande, 1749-016 Lisboa, Portugal}

\begin{abstract}
The star formation rate is a crucial astrophysical tracer for understanding the formation and evolution of galaxies, determining the interaction between interstellar medium properties and star formation, thereby inferring the evolutionary laws of cosmic star formation history and cosmic energy density. The mainstream approach to studying the stellar property in galaxies relies on pure stellar population synthesis models. However, these methods fail to account for the contamination of SFR caused by nebular gas radiation. Recent studies have indicated that neglecting nebular radiation contamination appears non-negligible in galaxies with intense star-forming activities and at relatively high redshifts, potentially leading to overestimating stellar masses. However, there is currently limited targeted research, particularly regarding galaxies at redshifts (z $<$ 0.4). In this work, 6,511 star-formation galaxies are selected from the SDSS-DR18, and FADO fits their spectra. This tool can exclude nebular radiation contributions in the spectral fitting. A tentative work is carried out to explore the SFR of these galaxies. The results indicate that the median \( H_{\alpha} \) flux obtained from FADO fitting differs from that obtained using the pure stellar population synthesis model {\it qsofitmore} by approximately 0.034 dex. Preliminary evidence suggests that the average nebula ratio increases with redshift. Additionally, we investigated the impact of stellar mass on the nebula ratio at low to moderate redshifts. By comparing two spectral fitting software packages, we found that although the contribution of nebular emission is minimal, it generally shows an increasing trend with redshift. We anticipate that by combining optical and near-infrared spectral data, the influence of nebulae may become more prominent in star-forming galaxies at higher redshifts (e.g., up to z $\sim$ 2).
\end{abstract}

\keywords{Galaxies -- Star-forming galaxies, Galaxies -- Spectra, Galaxies -- Star formation rate, Nebulae -- continuum, Techniques -- Spectroscopic, Code -- Python/Linux}

\section{Introduction} \label{sec:intro}
The star formation rate (SFR), particularly the instantaneous SFR (SFR(t)), serves as a significant tracer for tracking stellar formation activities in galaxies, understanding their evolutionary history, and elucidating galaxy chemical evolution \citep{Zhuang2019}. By combining observations of SFR with galaxy evolution simulations, the dominant physical processes driving galaxy growth can be thoroughly understood. For instance, processes such as galaxy quenching, gas accretion, star formation, and feedback from supernovae and active galactic nuclei (AGN) \citep{Krumholz2012, Mullaney2012}. The logarithmic linear relationship between stellar mass and SFR, known as the star-forming main sequence  \citep[SFMS, e.g., ][]{Noeske2007}, aids in understanding the trajectory of stellar activities in the Hertzsprung-Russell (HR) diagram and analyzing the characteristics of stars during the main sequence phase \citep{Tacconi2020}. By measuring SFR at different epochs in cosmic history, the contribution of various types of galaxies to the cosmic background radiation can be estimated, thereby refining cosmological ionization models and searching for the Pop III in the universe \citep{Madau2014}. To date, numerous efforts have been put into the SFR, for instance, variations in SFR during the formation of star clusters and massive stars in different environments\citep{Lada2003, McKee2002, Kirk2015, McKee2007}; the promotion of SFR or black hole growth due to interactions between galaxies \citep{DiMatteo2005, Springel2005}; SFR corrections for cosmological reionization \citep{Madau2014, WangFY2013, Robertson2012, Daigne2004}; reevaluation of star formation within individual molecular clouds and gas relations \citep{Kennicutt2012}. Several surveys in astrophysics involved studying the properties of star-forming galaxies, such as the Sloan Digital Sky Survey \citep[SDSS, ][]{York2000}, the Hubble Ultra Deep Field \citep[HUDF, ][]{Ellis2013}, and the Atacama Large Millimeter Array \citep[ALMA,][]{Wang2013}.

Mainstream methods for estimating the SFR typically involve a fundamental step of spectral fitting for galaxies. This process entails fitting the spectrum as a combination of pure stellar population components, utilizing models based solely on stellar populations as templates for galaxy spectral fitting (e.g., STARLIGHT \citep{CidFernandes2005}, pPXF \citep{Cappellari2004}, {\it qsofitmore} \citep{Fu2021}) Considering that SDSS galaxies exhibit minimal nebular radiation due to insignificant star-forming activities, the approaches above overlook the modelling of nebular gas \citep{Miranda2023}, thereby neglecting the influence of nebular emission \citep{Miranda2023, Ahumada2020}. The limitation of these pure stellar models lies in their inability to solve different processes of gas emission within galaxies, as they incorporate assumptions that combine stellar mass, gas mass, and dust mass into a singular framework. However, studies by \citet{Krueger1995} and \citet{Salim2007} indicate that vigorous star-forming activities can lead to the nebular continuum contributing approximately 30\%-70\% to optical and near-infrared emissions, with nebular emission around star-forming regions accounting for about 60\% of the total emissions \citep{Schaerer2009, Krumholz2012}. Therefore, when the acquired galaxy spectrum contains star-forming regions within the galaxy, the influence of nebulae becomes more pronounced. \citet{Pacifici2015} demonstrated that neglecting nebular emission in spectral modelling could lead to an overestimated SFR by approximately 0.12 dex. Ignoring the contribution of the nebular continuum would result in larger stellar mass and shallower ultraviolet slopes \citep{Izotov2024}. Galaxy evolution is primarily driven by transforming interstellar gas into stars, gas accretion between galaxies, intergalactic gas collisions, and galaxy mergers \citep{Scoville2023}. As the first independent and self-consistent spectral fitting tool to simultaneously consider stellar and nebular components in galaxies, FADO accounts for the impact of nebulae on SFR, dynamically and synchronously decomposing and calculating the contributions of different components and radiation mechanisms within the entire galaxy spectrum \citep{Gomes2017}. Therefore, we employ FADO as a research tool to consider nebular emission in this investigation.
    
In this work, we selected spectra of galaxies with active star formation at middle redshifts (z $<$ 0.4) from SDSS-DR18. These spectra were fitted using the FADO software and the Python toolkit {\it qsofitmore}, based on stellar population synthesis models. The aim was to explore whether nebular emission would interfere with the intrinsic estimation of SFR in galaxies at higher redshifts. $H_{\alpha}$ and $H_{\beta}$ are commonly regarded as the most reliable estimators of SFR \citep{Sobral2013}. However, more extensive consideration of the [O II]$\lambda$3728 line is required around z $\sim$ 0.5, which can serve as an estimator of SFR up to z $<$ 1.4 \citep{Kewley2004}. SFR below z $<$ 0.5 can be accurately traced by $H_{\alpha}$ \citep{Kennicutt1998}, moderate redshift SFR by [O II]$\lambda$3728 \citep{Kewley2004}, and high redshift SFR by [O III]$\lambda$5007 \citep{Shapley2023}. 
    
Several studies have recently demonstrated the stability of FADO in spectral fitting and line information estimation. \citet{Breda2022} found, through comparison with STARLIGHT using simulated data, STARLIGHT overestimated the mass-weighted mean stellar age, light-weighted mean stellar age, and metallicity. These findings concluded that FADO can accurately recover mass, age, and mean metallicity with high precision ($\sim$ 0.2 dex). Furthermore, \citet{Cardoso2022} showed that for galaxies with EW (H$_{\alpha}$) \textgreater 3 $\AA$, the average mass and metallicity of the entire galaxy can differ by up to $\sim$ 0.06 dex. Most recently, \citet{Miranda2023} compared the SFR obtained from FADO fitting of the MPA-JHU galaxy sample with that obtained from STARLIGHT fitting of the MPA-JHU sample and found a difference of 0.01 dex in Ha flux between the two methods. They also predicted that nebular radiation not only plays a crucial role in the star formation activity of galaxies but also affects the calculation of galaxy SFR, particularly in high-redshift galaxies. Additionally, recent research by \citet{Izotov2024} has shown that neglecting the contribution of the nebular continuum in determining SFR would result in larger stellar masses and shallower ultraviolet slopes. This is particularly significant for galaxies like J2229+2725, which are considered analogues of early-universe dwarf galaxies and play a role in cosmic reionization.
    
This work aims to compare the spectroscopic fitting of the pure stellar code {\it qsofitmore} and FADO to explore the possible impact of the nebular contribution on SFR estimation. And explore the relationship between this effect and redshift. By examining the differences in the $H_{\alpha}$ emission line in the fitted spectra, we aim to further determine the disparity in SFR. It is worth mentioning that based on the rough information available about stars in galaxies, stellar mass and SFR may not be sufficient to infer the potential evolutionary paths of galaxies. We still need to combine the spatial resolution information provided by all baryonic components of galaxies (stars, gas, metals) and the complete kinematic tracers of gravitational potential (dark matter), as well as the kinematic tracers of galaxy feedback processes (quenching, AGN, and supernova feedback), to obtain the complete evolutionary paths of galaxies \citep{Forster2020}. Therefore, we remain open-minded regarding the derivation of galaxy evolution processes in the current scenario.
    
\hyperref[Section2]{Section 2} outlines the criteria for selecting sources in the galaxy spectroscopic analysis and the process of spectral fitting employed in this study. In \hyperref[Section3]{Section 3}, we present the statistical analysis of the selected sample and the main findings. \hyperref[section4]{Section 4} discusses four critical issues related to the spectroscopic data and fitting techniques employed in this investigation. Finally, \hyperref[section5]{Section 5} summarizes this preliminary research. Throughout the paper, we assume a standard cosmological model with Hubble constant $H_0$ = 72 km s$^{-1}$ Mpc$^{-1}$, matter density $\Omega_M$ = 0.3, and cosmological constant $\Omega_{\Lambda}$ = 0.7.

\section{sample selection and spectral fitting} \label{Section2}
The galaxy sample and their spectra used in this work are selected from Sloan Digital Sky Survey Data Release 18th \citep[SDSS-DR18;][]{Almeida2023}. It integrates data from the previous 17 data releases and incorporates new observations and improvements. The galaxy sample was obtained by SQL query, and their spectroscopic data were retrieved from the SDSS Science Archive Server (SAS) \footnote{\url{https://dr18.sdss.org/optical/spectrum/search}}.

\subsection{SQL query} \label{Subsection2.1}
To guarantee the subsequent spectral fitting, it is necessary to select spectra with sufficient signal-to-noise ratio (S/N). Since FADO performs better with median S/N of spectra larger than 5 \citep{Pappalardo2021P}, the SQL query was adopted to select high S/N spectra. The query obtained 91,252 spectra with 5 \textless\ S/N \textless\ 20 for 0 \textless\ z \textless\ 0.2 galaxies, and 6,604 with S/N \textgreater\ 5 for 0.2 \textless\ z \textless\ 0.4 galaxies. Additionally, the S/N of major emission lines of each spectrum was also constrained by SQL query with S/N \textgreater\ 5. The following conditions were added in the query to ensure the reliability of redshift measurements: \texttt{zwarning = 0}, \texttt{veldisperr > 0}, \texttt{veldisp > 3 * veldisperr}, and \texttt{80 < veldisp < 350} (see \autoref{AppendixA} for detailed SQL query).

\subsection{BPT selection} \label{Subsection2.2}
\begin{figure}[H]
    \centering
    \includegraphics[width=1\linewidth]{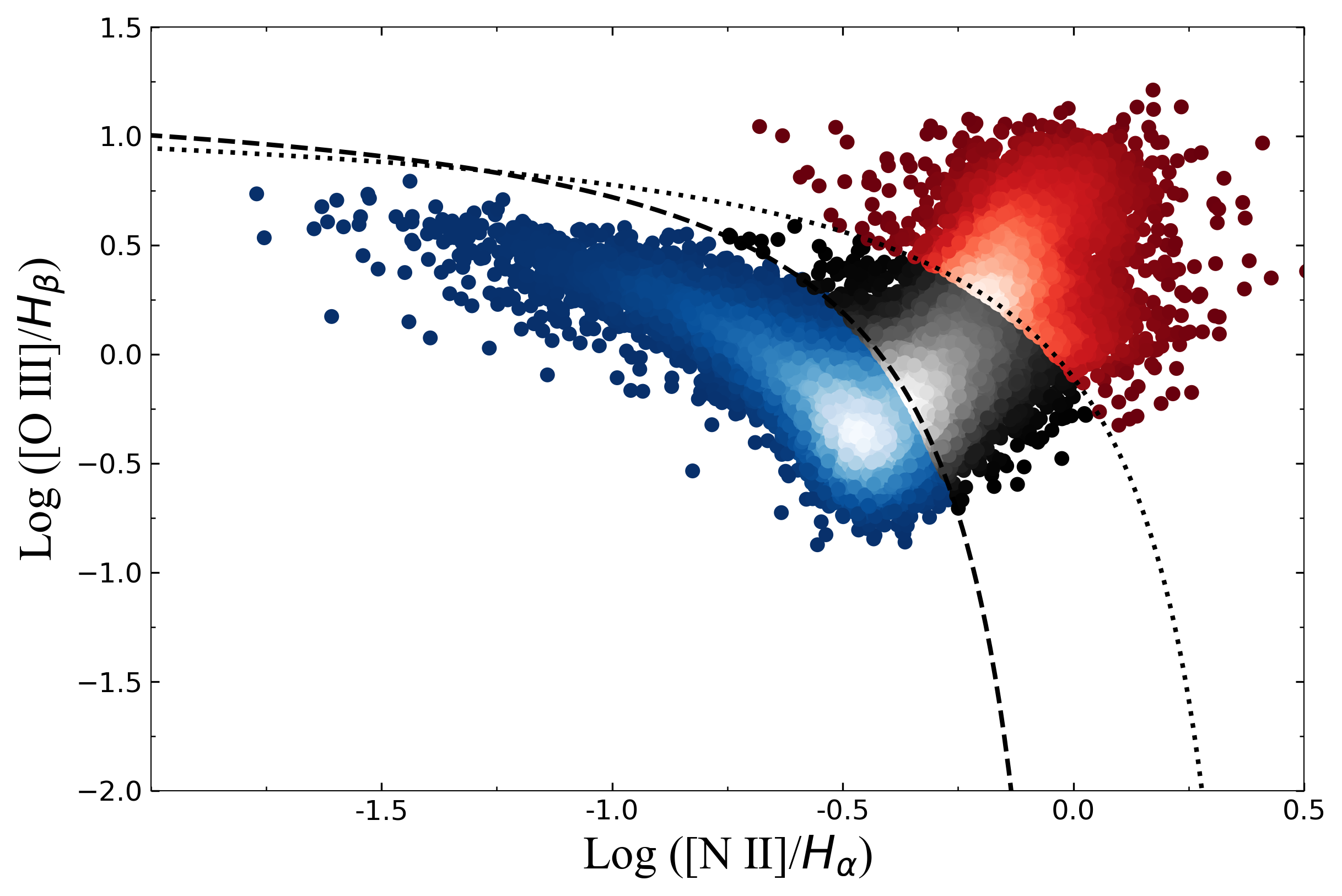}\vspace{-6pt}
    \caption{The distribution of the 21,455 galaxies on BPT diagram. The blue, black, and red scatters represent star-forming, composite, and AGN/Seyfert galaxies. The denser the scatter, the whiter the colour.}
    \label{fig1}
\end{figure}

After SQL selection, the BPT diagram \citep{Baldwin1981} was adopted to isolate the SFGs from the sample. The two line ratios, [N II]$\lambda$6583/$H_{\alpha}$ and [O III]$\lambda$5007/$H_{\beta}$, were selected to build the BPT diagram. Two boundaries divide the sample into three regions: star-forming, composite, and AGN. In \hyperref[fig1]{Figure 1}, the dashed boundary \hyperref[eq1]{Equation (1)} is derived from \citet{Kewley2001}, while the dotted one \hyperref[eq2]{Equation (2)} from \citet{Kauffmann2003}. The samples in the pure star-forming region of the BPT diagram were selected as the SFGs (blue scatter in \hyperref[fig1]{Figure 1}).

\vspace{-6pt}

\begin{equation}
   \log \left(\frac{[\mathrm{O} \mathrm{III}] \lambda 5007}{\mathrm{H}_{\beta}}\right)=\frac{0.61}{\log \left([\mathrm{N} \mathrm{II}] \lambda 6538 / \mathrm{H}_{\alpha}\right)-0.47}+1.19,
\end{equation}\label{eq1}

\begin{equation}
   \log \left( \frac{[\mathrm{O} \mathrm{III}] \lambda 5007}{\mathrm{H}_{\beta}} \right) > \frac{0.61}{\log([\mathrm{N} \mathrm{II}] \lambda 6538 / \mathrm{H}_{\alpha}) - 0.05} + 1.3.
\end{equation}\label{eq2}

\begin{figure}[H]
    \centering
    \includegraphics[width=1\linewidth]{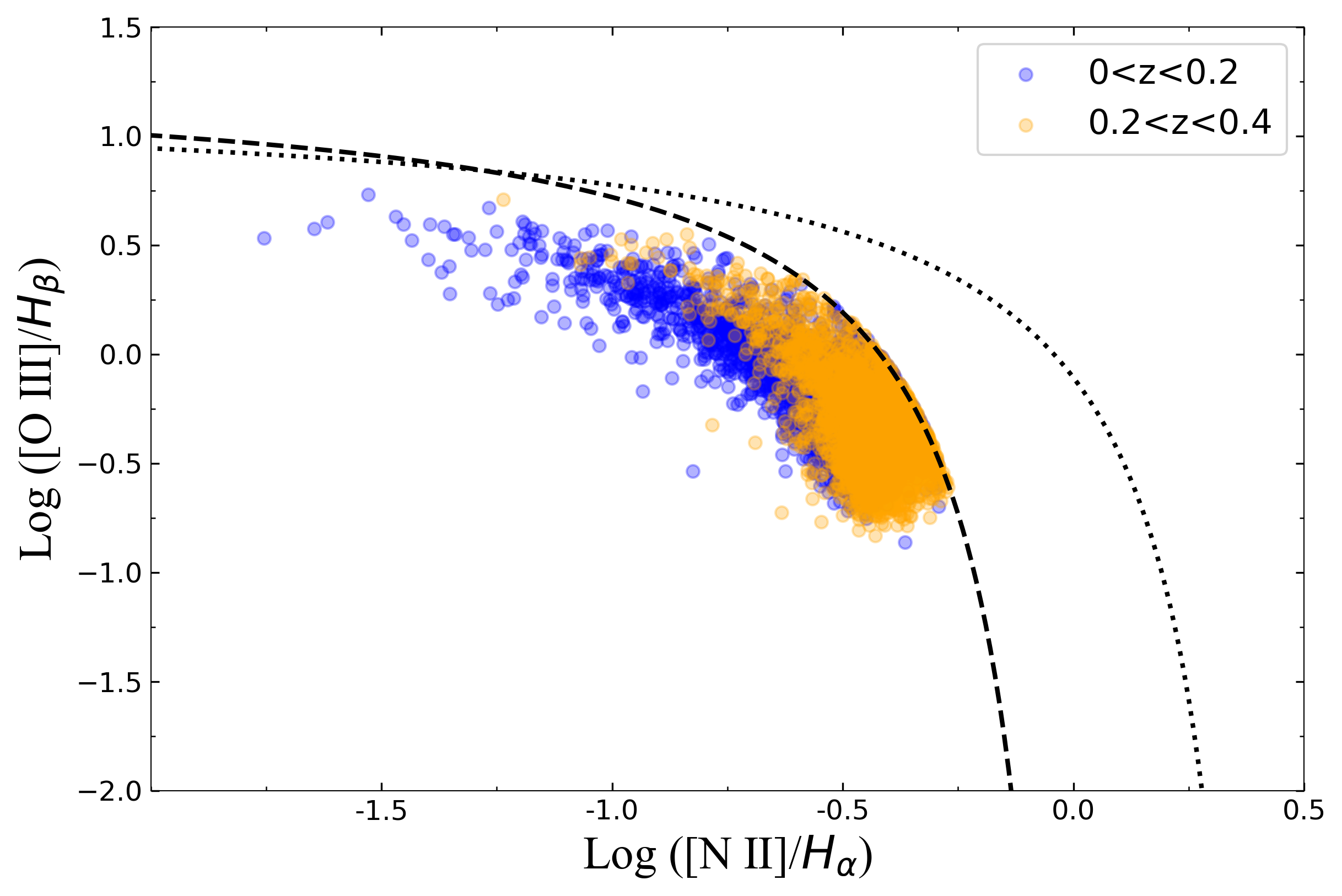}\vspace{-6pt}
    \caption{The distribution of the lower-redshift (blue) and higher-redshift (orange) samples. The dashed and dotted boundaries are the same as \hyperref[fig1]{Figure 1}.}
    \label{fig2}
\end{figure}

A total of 21,455 galaxies were considered SFGs, which is a composite of 18,176 galaxies at 0 \textless\ z \textless\ 0.2 (lower-redshift sample) and 3,279 galaxies at 0.2 \textless\ z \textless\ 0.4 (higher-redshift sample). To avoid the bias of the larger number of lower-redshift galaxies, 3,279 of them were randomly selected to stand for the lower-redshift sample. Ultimately, a sample containing 6,558 galaxies was submitted to fit by {\it qsofitmore} and FADO, aiming at deriving more accurate emission line fluxes. \hyperref[fig2]{Figure 2} shows the distribution of the two samples with different redshift intervals.

\vspace{-6pt}

\subsection{Spectral fitting} \label{Subsection2.3}

The {\it qsofitmore} is a Python module dedicated to fitting and analyzing spectra of quasars and galaxies \citep{Fu2021}. The module has shown good performance in spectral fitting and analysis and has become increasingly popular in quasar and galaxy spectral studies \citep{Fu2022, Jin2023, Ding2023, Chen2024}. The {\it qsofitmore} not only fits the emission lines of the spectra but also uses the Monte Carlo (MC) method to obtain the uncertainties of the fitted components. It should be noted that the module is based on the BC03 model \citep{Bruzual2003} for spectral fitting, which does not simultaneously consider the stellar and nebular radiations to decompose the spectral components, as FADO does. This allows us to compare the $H_{\alpha}$ obtained by FADO and {\it qsofitmore} to demonstrate whether a significant difference exists.

FADO is a self-consistent spectral fitting tool that can dynamically calculate the parameters of a galaxy spectrum. It estimates the contribution of nebular radiation from emission line luminosity. FADO estimates nebular emission lines through the Full-Consistency, Nebular-Continuum, and STellar mode \citep{Gomes2017}. The main difference between FADO and other pure stellar codes (such as STARLIGHT and {\it qsofitmore}) is the consideration of nebular emission or not. Both {\it qsofitmore} and FADO adopt basic template spectra derived from the BC03 model and estimate parameter uncertainties with the MC process. In addition, both of them can perform extinction correction, de-redshift, and K-correction on spectra under given coordinates and redshifts. These may ensure the comparability of spectral fitting. A comparison of the spectral fitting by {\it qsofitmore} and FADO for the same galaxy is displayed in \autoref{AppendixB}.

\vspace{-6pt}
\begin{figure}[H]
    \centering
    \includegraphics[width=1\linewidth]{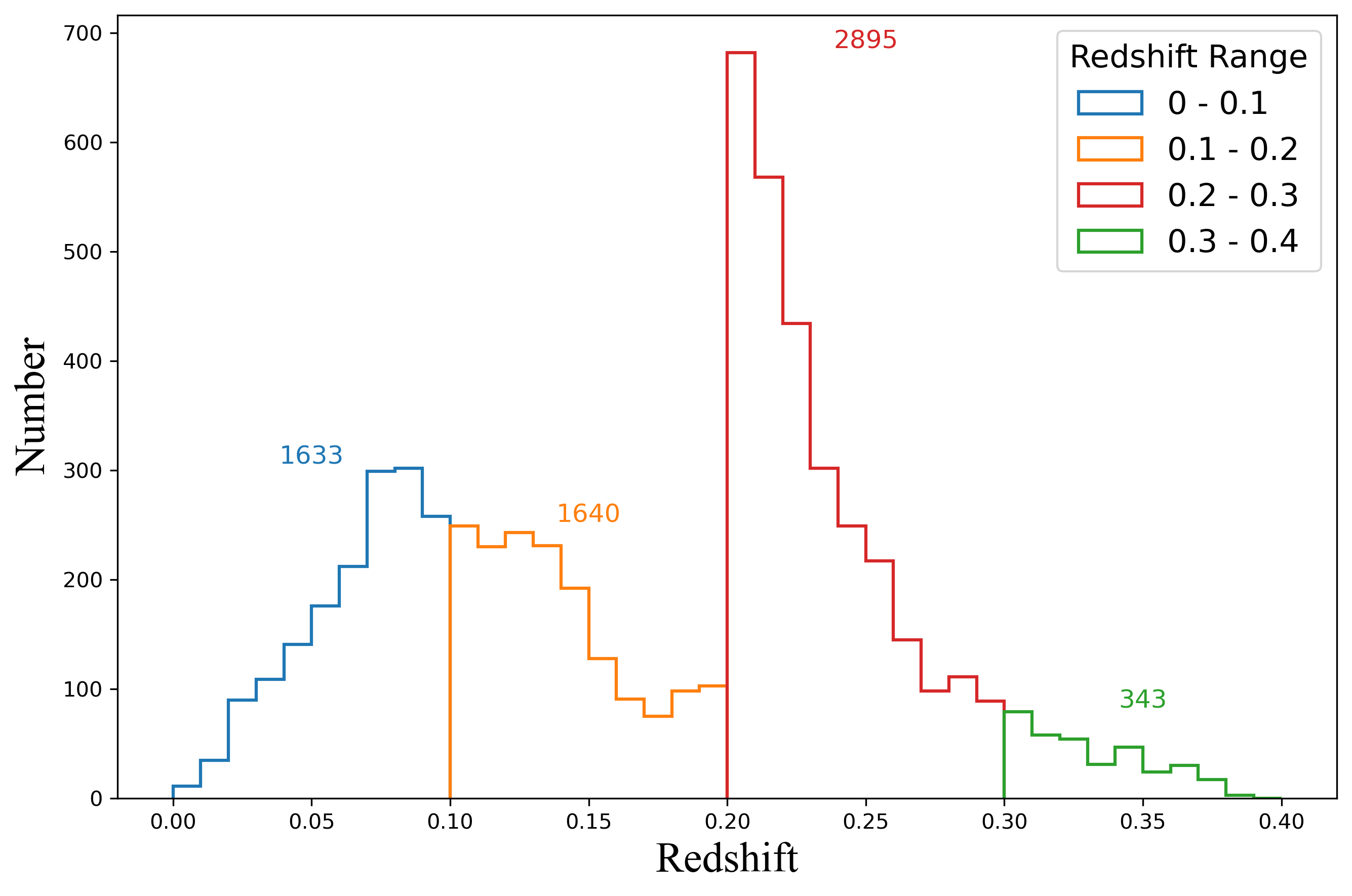}\vspace{-6pt}
    \caption{The statistics of the redshift and quantity of the \texttt{Golden Sample}. The redshift is binned with the interval of 0.1 to show in detail the scarcity of the SDSS galaxy spectrum at higher redshift.}
    \label{fig3}
\end{figure}

Following the comparison above between the two fitting tools, we conducted spectral fitting on the 6558 galaxies obtained in \hyperref[Subsection2.1]{Section 2.1} using {\it qsofitmore} and FADO, respectively. To consider intrinsic extinction correction, the two fittings must give physically meaningful results of the flux of $H_{\alpha}$ and $H_{\beta}$, as well as their corresponding uncertainties. Then 6,546 and 6,523 galaxies were filtered out by {\it qsofitmore} and FADO, respectively. Finally, a total of 6,511 galaxies were selected as the \texttt{Golden Sample}, in which the spectra of galaxies are successfully fitted by both FADO and {\it qsofitmore}. Their statistics of the redshift and quantity are displayed in \hyperref[fig3]{Figure 3}. Notably, there is an obvious scarcity of quantity in the redshift interval from 0.3 to 0.4. This is due to the deficiency of the SDSS survey on the higher-redshift galaxy. The redshift range of 0.3 $\sim$ 0.4 exceeds the median redshift of the SDSS survey \citep[around 0.1, ][]{Abazajian2009}, leading to a noticeable drop in the sample quantity. Besides, $H_{\alpha}$ may fall beyond the maximum wavelength coverage of the SDSS around a redshift of 0.4, resulting in its failure to be fitted. Therefore, the requirement of the BPT diagnosis constraints more strictly on sample volume, particularly in the redshift range of 0.3 $\sim$ 0.4. Although galaxies in this range are relatively sparse, these samples are important and cannot be ignored in this work.

\subsection{Intrinsic extinction correction} \label{Subsection2.4}
There are sufficient young and hot stars in the star-forming regions that provide enough energy for the ionization of hydrogen. The ionized gas captures photons and undergoes recombination transitions, leading to simultaneous emission of $H_{\alpha}$ and $H_{\beta}$. However, cold gas and dust heavily attenuate the significant electromagnetic radiation emitted during star formation. Since attenuation of cold matter is more efficient at shorter wavelengths than longer ones, this results in a decrease in the $H_{\alpha}$-$H_{\beta}$ ratio \citep{Calzetti2000}. Commonly, a $H_{\alpha}$-$H_{\beta}$ ratio of 2.86 is regarded as a typical intrinsic extinction corrector \citep{Gaskell1984, Hummer1987, Cardelli1989}. Therefore, the intrinsic extinction of \texttt{Golden Sample} was corrected with the ratio.

\section{analysis and results} \label{Section3}
This section compares the $H_{\alpha}$ flux of \texttt{Golden Sample} fitted by both {\it qsofitmore} and FADO to assess whether the two fittings are consistent. The nebular contribution in the wavelength range of $H_{\alpha}$ is calculated to examine the variation with redshift. The SFR is also estimated with the $H_{\alpha}$ flux fitted by FADO to evaluate whether FADO can provide a more reasonable SFR by excluding nebular contamination.

\subsection{Comparison of \texorpdfstring{\( H_{\alpha} \)}{H-alpha} flux}\label{Subsection3.1}
\begin{figure}[H]
    \centering
    \includegraphics[width=1\linewidth]{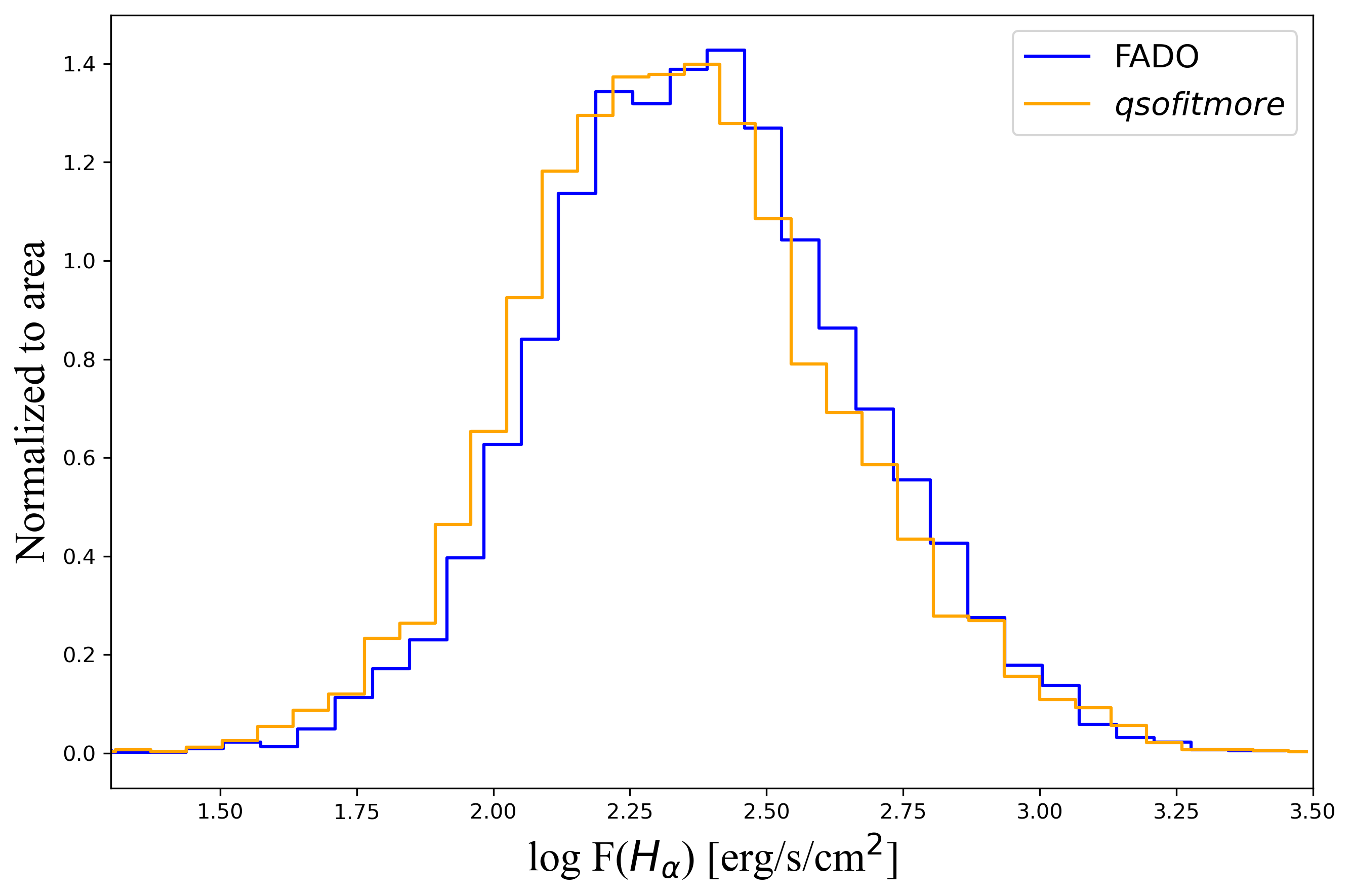}\vspace{-6pt}
    \caption{The distribution of $H_{\alpha}$ flux fitted by {\it qsofitmore} (orange histogram) and FADO (blue histogram). The $H_{\alpha}$ flux is logarithmized, and the number of them is normalized.}
    \label{fig4}
\end{figure}

\begin{figure}[H]
    \centering
    \includegraphics[width=1\linewidth]{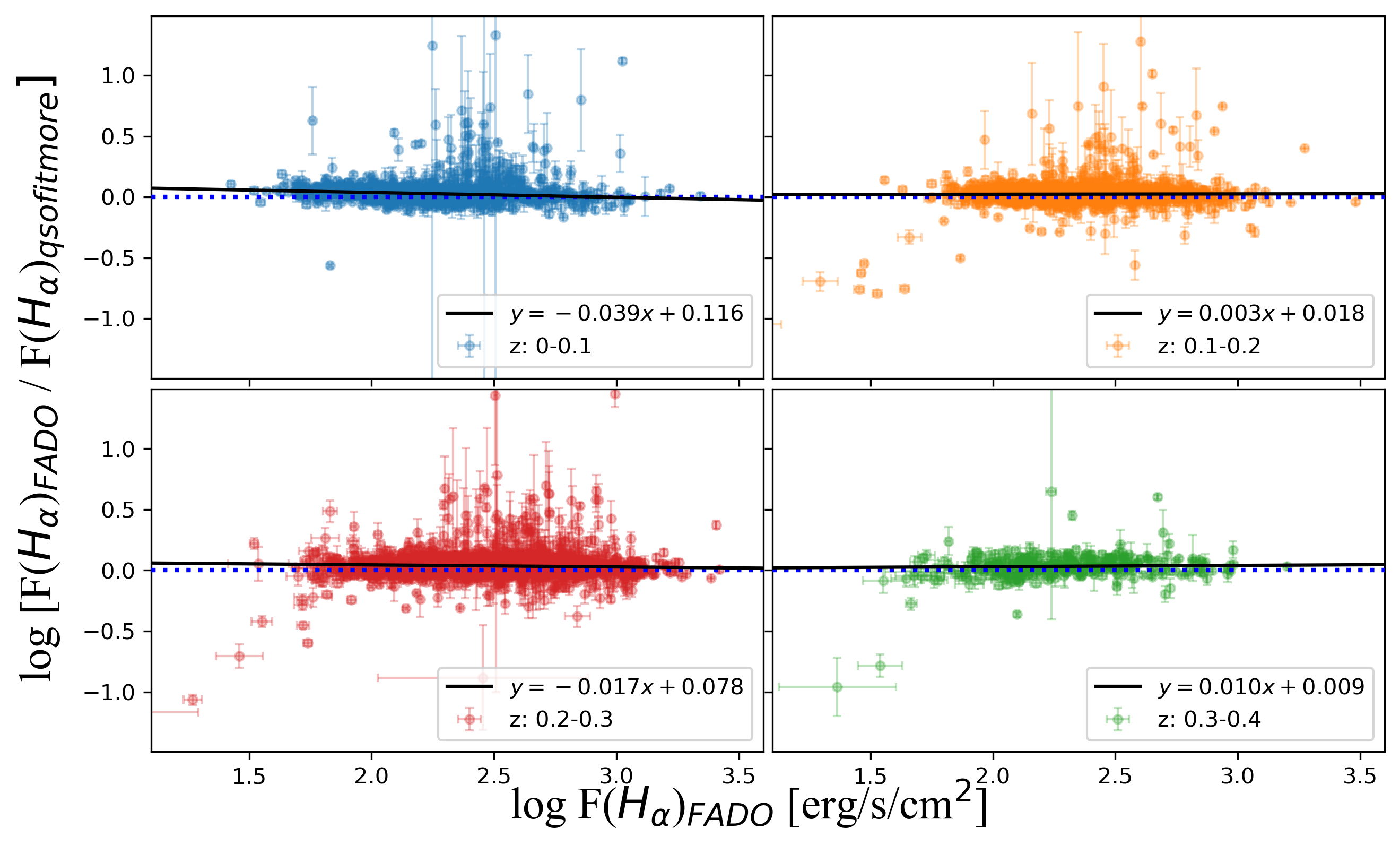}\vspace{-6pt}
    \caption{The redshift binned logarithmized $H_{\alpha}$ flux ratio versus the logarithmized $H_{\alpha}$ flux fitted by FADO. The flux ratio in each panel is calculated by FADO fitted and {\it qsofitmore} fitted $H_{\alpha}$. In each panel, the blue dotted lines represent the equality of the $H_{\alpha}$ flux fitted by {\it qsofitmore} and FADO. In contrast, the black solid lines denote the linear fitting of the scatter considering uncertainties.}
    \label{fig5}
\end{figure}

As is shown in \hyperref[fig4]{Figure 4}, the $H_{\alpha}$ flux obtained from FADO is statistically similar to that from \textit{qsofitmore}, which is consistent with the result in Figure 2 of \citet{Miranda2023}. KS test was performed on the \textit{qsofitmore}, and FADO fitted $H_{\alpha}$ flux to assess whether they follow the same distribution. The KS statistic is 0.068, and the p-value is $1.68 \times 10^{-13}$, suggesting statistical uniformity between the two fitted $H_{\alpha}$ subsets. Further, the $H_{\alpha}$ flux was binned by redshift with the interval of 0.1 to explore whether there exists a tendency with redshift between the two fittings. The results in \hyperref[fig5]{Figure 5} indicate no clear hint of $H_{\alpha}$ difference caused by redshift up to at least z $\sim$ 0.4. In general, the $H_{\alpha}$ flux fitted by both {\it qsofitmore} and FADO do not exhibit noticeable differences, which is consistent with the results of \citet{Miranda2023}. The Middle Difference is \( 0.034 \, \text{dex} \approx 8.16\% \).

\vspace{-6pt}

\vspace{-6pt}

\subsection{Nebular Contribution}\label{Subsection3.2}

\begin{figure*}[!t]
\centering  
\subfigure{
\label{Fig.sub.1}
\includegraphics[width=0.48\textwidth]{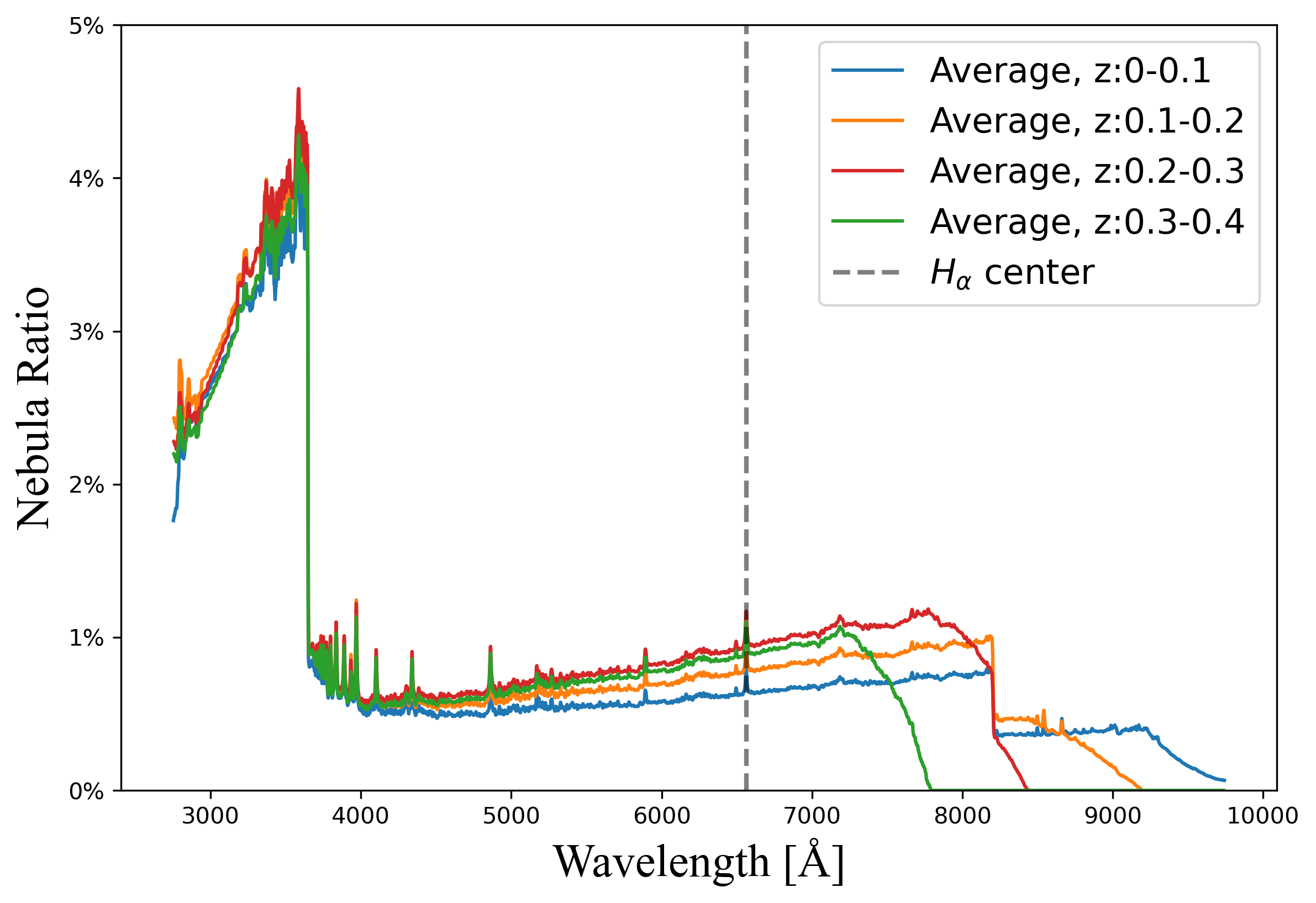}}
\subfigure{
\label{Fig.sub.2}
\includegraphics[width=0.48\textwidth]{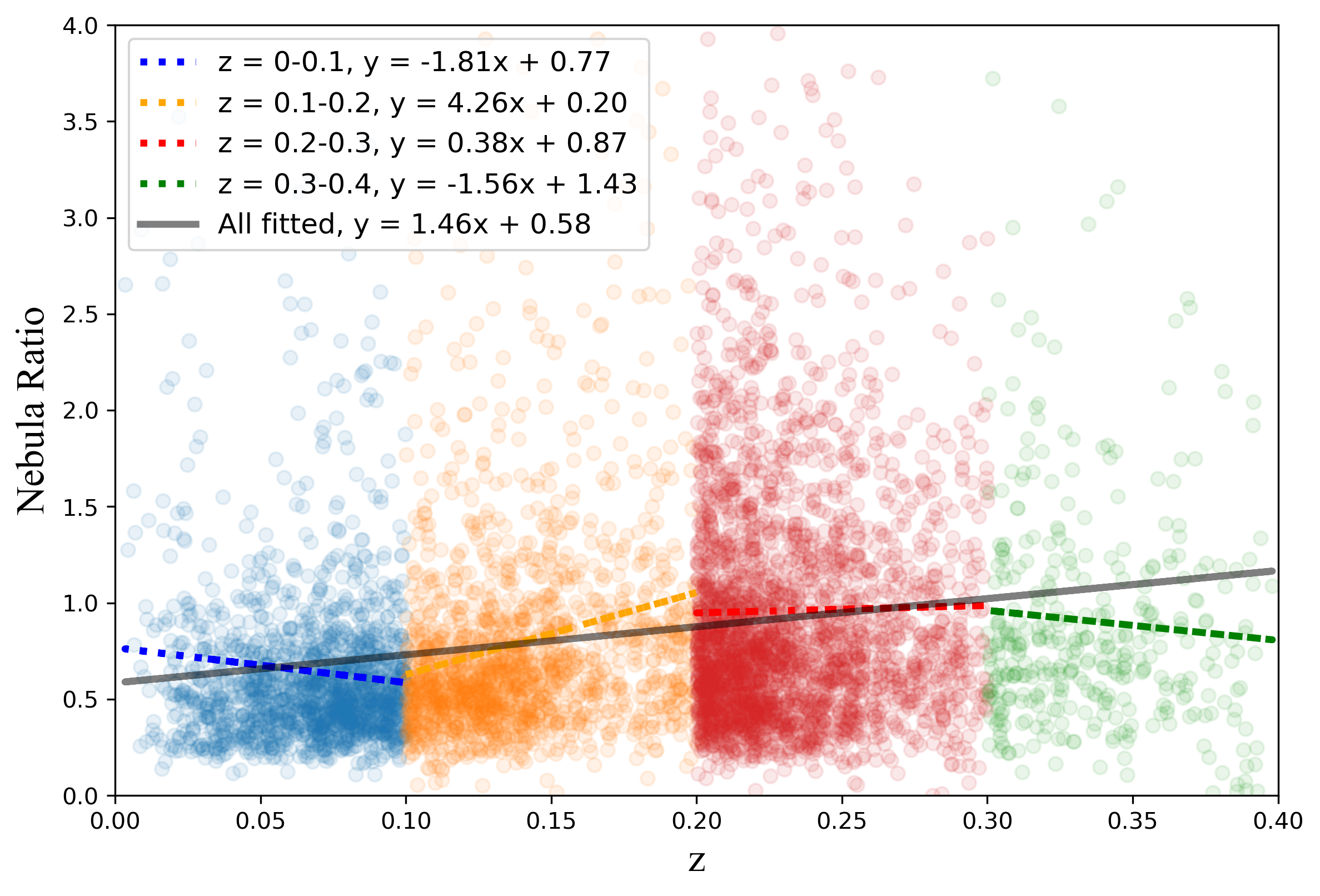}}
\caption{\textit{Left panel}: Fitted nebula components. The average nebula spectra in different redshift ranges are shown, and $H_{\alpha}$ centre indicates how the average nebula spectrum at $H_{\alpha}$ wavelength changes with redshift. \textit{Right panel}: Nebular contribution at $H_{\alpha}$. The nebular radiation of all regions of $H_{\alpha}$ is added together. The five trend lines represent the trend of the data at their respective redshifts and the overall trend.}
\label{fig6}
\end{figure*}

To further analyze and quantify the possible nebular contribution, the fitted nebular spectra were extracted from the FADO fitting outputs. The fitted nebula components were binned by redshift and averaged within the corresponding redshift intervals. These averaged nebular spectra displayed in panel (a) of \hyperref[fig6]{Figure 6} increase with the increase of redshift at the centre wavelength of $H_{\alpha}$ (black dashed line), except for the averaged spectra in the interval of 0.3 \textless\ z \textless\ 0.4 (green line). The decline of the green averaged spectra relative to the red one would be due to the lack of the sample and the low quality of the spectra. We expect the green averaged spectra to be raised after the expansion of the higher-redshift sample. The shape of the nebula radiation in \hyperref[fig6]{Figure 6} panel (a) is all uniform, it comes from the FADO fitting template. For details, refer to \citet{Gomes2017}.

\hyperref[fig6]{Figure 6} panel (b) shows that the nebula ratio of the fitted nebular spectra was also integrated within the wavelength range from 6,400 \AA\ to 6,800 \AA\ to quantify the impact on $H_{\alpha}$. The Nebula ratio is equal to the Nebula Model fitted in FADO divided by the Stellar Model. This method visualizes the percentage of nebulae in the entire spectrum. For a detailed explanation, please see \autoref{AppendixB}. \hyperref[fig6]{Figure 6} panel (b) presents the linear analytic expressions corresponding to the four redshift intervals, plotted in the colours of their respective scatter points, along with an overall trend represented by a black solid line. Calculating the cumulative nebular contribution at wavelengths between 6,400 \AA\ and 6,800 \AA\ for the samples in different redshift intervals reveals a proportional increase from $z=0.1-0.2$ to $z=0.2-0.3$. This indicates that for the samples used in this work, the nebular contribution increases, in fact, with the redshift, directly confirming the hypothesis proposed by \citet{Miranda2023}.

However, the increase in redshift is not stable and has a large dispersion. The Pearson correlation coefficient can be used to better quantify this issue. The Pearson correlation coefficient for $z=0-0.1$ is -0.077, for $z=0.1-0.2$ is 0.189, for $z=0.2-0.3$ is 0.013, and for $z=0.3-0.4$ it is -0.065. The total Pearson correlation coefficient for the four redshift bins is 0.181. Overall, the nebula ratio has a more obvious trend with redshift, with a slope of 1.46. The reason for the difference in the sensitivity of the nebula ratio to redshift in the above four redshift bins may be the instability of gases (especially the density of hydrogen) at medium and low redshifts.

The hydrogen density in galaxies is also related to the epoch of the galaxy. Early-type galaxies contain far more gas than late-type galaxies \citep{rhee2013}. Since this work uses the H\(\alpha\) emission line to compare the differences between two fitting codes and utilizes the H\(\alpha\) flux to estimate the SFR, the hydrogen density value has a significant impact on the nebula ratio. Figure 9 of \citet{rhee2013} summarizes the effect of redshift on hydrogen density. The H\(_2\) density at \( z = 0 \) is defined by \citep{zwaan2005}, at \( z = 0.1 \) by \citep{martin2010}, at \( z = 0.2 \) by \citep{delhaize2013}, around \( z = 0.3 \) by \citep{freudling2010}, and around \( z = 0.4 \) by \citep{rhee2013}. Therefore, the H\(_2\) density at \( z = 0.4 \) should be the lowest, which corresponds to the trend in panel (b) \hyperref[fig6]{Figure 6}, where the nebula ratio decreases with redshift in \( z = 0.4 \). At \( z = 0.1 \), due to the depletion of cold gas, the galaxy evolution becomes static, which leads to a decreasing trend in the nebula ratio.

For galaxies with stellar masses greater than \( 10 \, M_{\odot} \), the content of H\(_2\) + He follows a trend of first increasing, then stabilizing, and eventually decreasing in the redshift range \( 0 < z < 0.4 \), as shown by \citep{geach2011,pavesi2018}. So, we would expect that the relationship between the nebular ratio and the redshift would yield more significant results if the stellar mass and galaxy type were consistent.

\subsection{Stellar Mass Effect}\label{Subsection3.3}
For the nebula ratio, there is still an invisible variable, the stellar mass of the galaxy, which affects the redshift relationship. This is due to the limitations of human sky photography techniques. This effect can be obtained from the z-SFMS diagram.

\begin{figure*}[!t]
\centering  
\subfigure[FADO's SFMS Figure]{
\label{Fig.sub.1}
\includegraphics[width=0.48\textwidth]{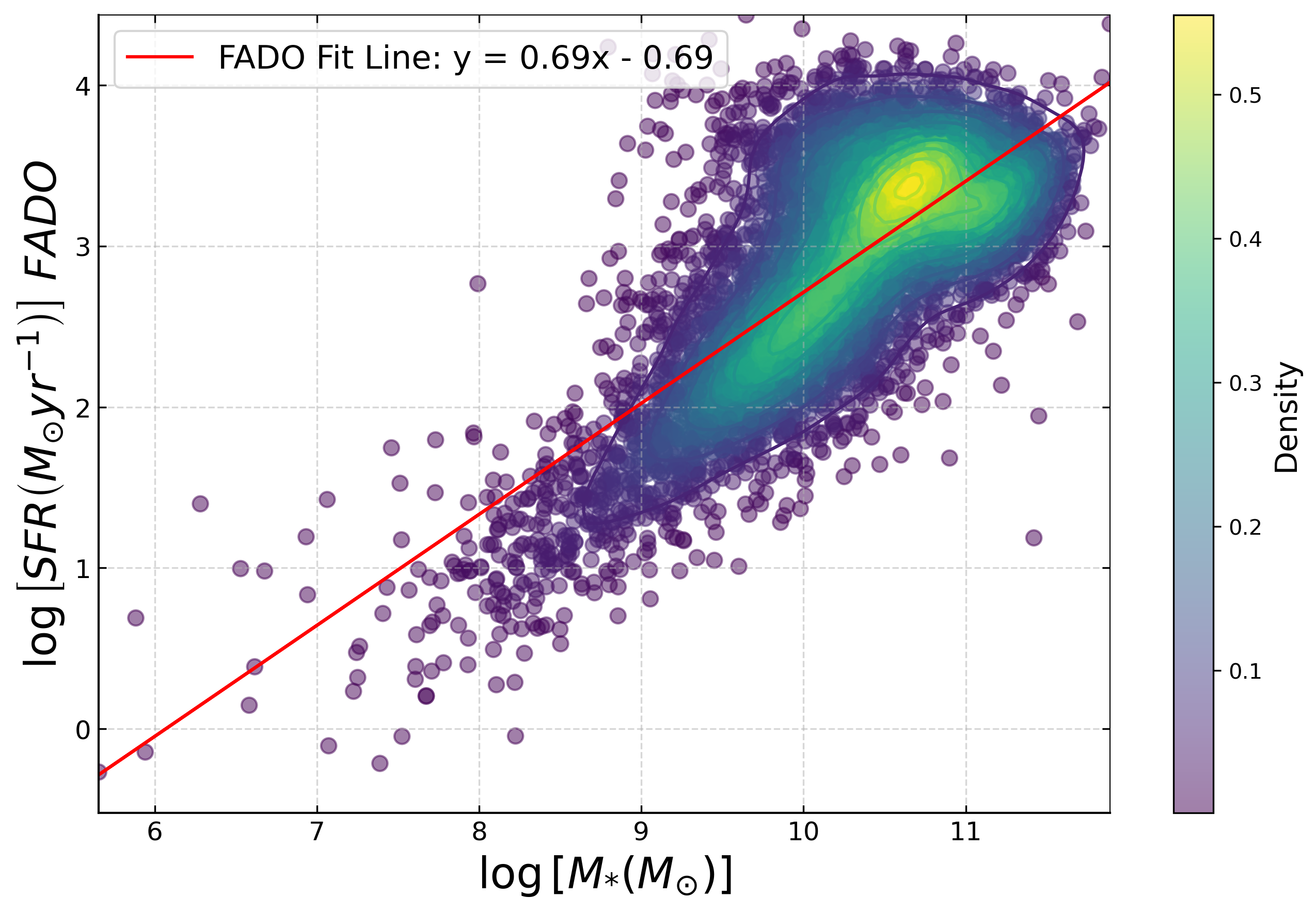}}
\subfigure[FADO's z-SFMS Figure]{
\label{Fig.sub.2}
\includegraphics[width=0.48\textwidth]{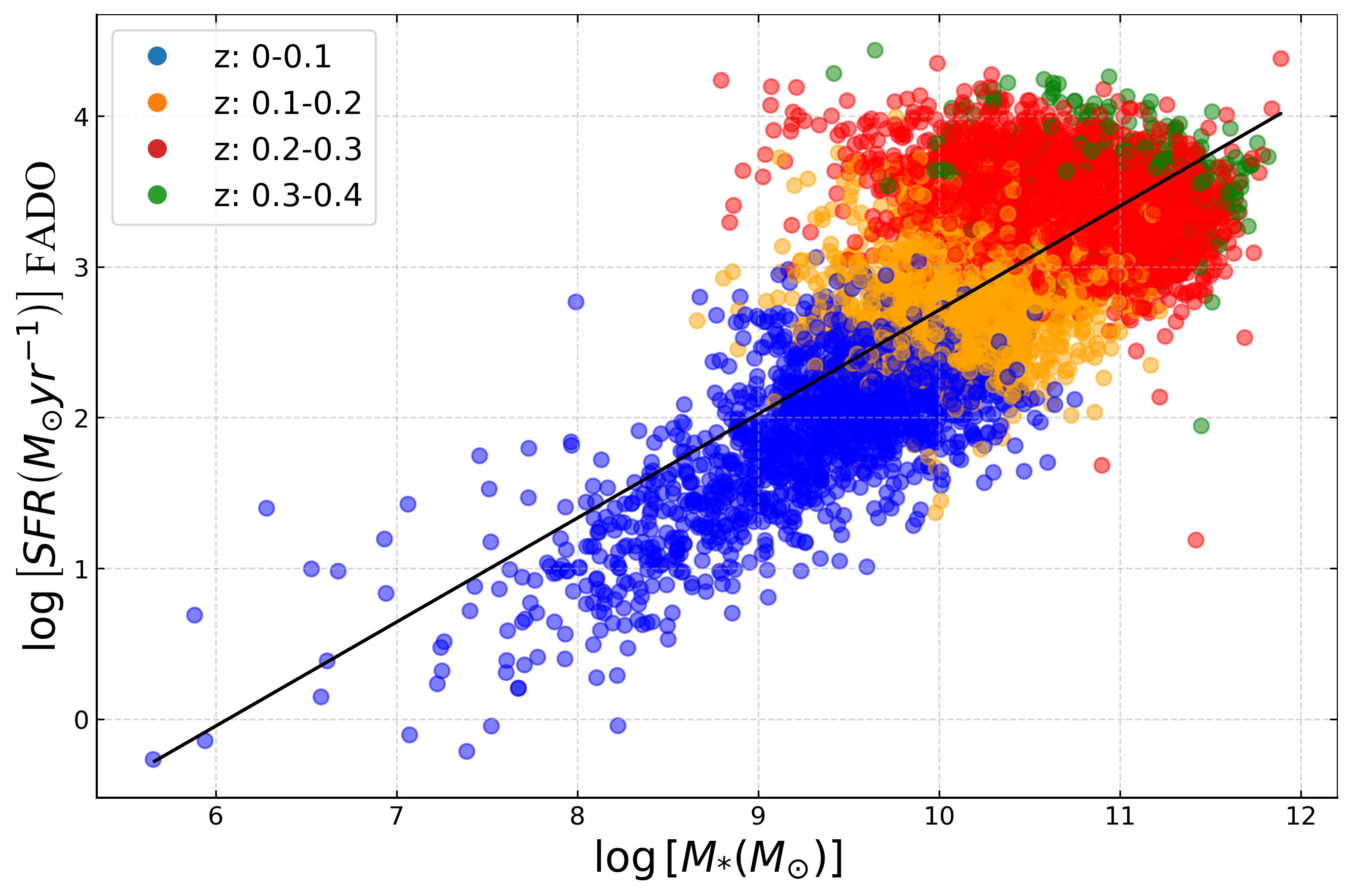}}
\caption{\textit{Left panel}: SFMS diagram fitted by FADO. The red line shows the slope of the SFR calculated by FADO fitting $H_{\alpha}$ flux and the stellar mass simulated by FADO. Contour lines indicate the data distribution: the denser the data points, the brighter the regions in the figure, and vice versa. \textit{Right panel}: z-SFMS diagram fitted by FADO. The black line represents the slope of the SFR calculated by FADO compared to the stellar mass, which is equal to the slope in the left panel. Different colours represent different redshift intervals.}
\label{fig7}
\end{figure*}

Despite its limitations, the $H_{\alpha}$ flux is still widely considered the most accurate estimator of galaxy star formation rate for nearby galaxies compared to other estimators, and it is the primary tracer used in this analysis. Therefore, this investigation utilizes the $H_{\alpha}$ luminosity to calculate the SFR of the golden sample. \hyperref[eq3]{Equation (3)} and \hyperref[eq4]{Equation (4)} are employed to compute the optimal SFR values.

\begin{equation}
    L(\mathrm{H}_\alpha) = F(\mathrm{H}_\alpha) 4 \pi d_L^{2},
\end{equation}\label{eq3}
\begin{equation}
    \mathrm{SFR} = \frac{L(\mathrm{H}_\alpha)}{\eta(\mathrm{H}_\alpha)}.
\end{equation}\label{eq4}
Where $d_L$ is the luminosity distance, $F\ (\mathrm{H}_\alpha)$ the best fitted flux of after extinction of $H_{\alpha}$, $L\ (\mathrm{H}_\alpha)$ the luminosity of $H_{\alpha}$ and $\eta(\mathrm{H}_\alpha)=10^{41.28} \mathrm{erg} \mathrm{s}^{-1} M_{\odot}^{-1} \mathrm{yr}$, the multiplication factor given by \citep{Miranda2023}.Calibrated to \citet{Chabrier2003} IMF. Among them, {\it qsofitmore} and FADO use consistent calibration and consistent calculation methods to calculate SFR.

Next, we attempt to quantify the differences in SFR calculations using FADO. Since {\it qsofitmore} does not have the capability to calculate stellar mass, only FADO data are used to plot the SFMS. The SFR is calculated using the equations as shown in \hyperref[eq3]{Equation (3)} and \hyperref[eq4]{Equation (4)}. Intrinsic extinction corrections for galaxies have been considered during the SFR calculations. A strong correlation between stellar mass and SFR is observed, with current studies indicating a positive log-linear relationship between them. As the SFR calculation is based on extinction-corrected data, reasonable values for $H_{\alpha}$ and $H_{\beta}$ must be obtained from the fits. The left panel of \hyperref[fig7]{Figure 7} shows the SFMS diagram derived from FADO, while the right panel shows the variation of the SFMS diagram with redshift. The red and black lines in \hyperref[fig7]{Figure 7} represent the slopes of the SFR calculated from the left and right panels, respectively. As they are derived from the same diagram, the slopes are consistent. The data from this investigation also conform to the established SFMS relationship from other studies, in which overall SFR increases with stellar mass, consistent with the findings of \citet{Miranda2023} in Figure 9. Ultimately, \hyperref[fig7]{Figure 7} shows that the relationship between stellar mass and SFR is described as $\log \left[\text{SFR}\left(M_{\odot} \, \text{yr}^{-1}\right)\right]_\text{FADO} = 0.69 \log \left[M_{*}\left(M_{\odot}\right)\right] - 0.69$.

In the right panel of \hyperref[fig7]{Figure 7}, it is evident that stellar mass and redshift exhibit a strong positive correlation, as does SFR with redshift. This clear positive relationship arises from the early universe (high redshift), where gas density and energy levels were higher, allowing gas to collapse more efficiently to form stars. In the later stages of the universe (mid-to-low redshift), as gas is depleted—excluding external factors such as gas replenishment or galaxy mergers—star formation rates decline accordingly. This decline begins around the epoch known as "cosmic noon" (z $\sim$ 2) \citep{Behroozi2013}. Across the redshift range 0-8, a positive linear relationship is observed between $\log M_*$ and $\log$SFR \citep{Madau2014, Behroozi2013}. Stellar formation density also increases with redshift in the ranges 0-0.8 and 0-5 \citep{Hopkins2006}, and stellar mass shows a strong positive correlation with redshift in the range 0-4 \citep{Ilbert2013}.

\begin{table}[ht]
\label{tab1}
\centering
\renewcommand{\arraystretch}{1}
\setlength{\tabcolsep}{6pt}
\caption{Analysis of the Stellar Mass and SFR in different redshift intervals.}
\begin{tabular}{l@{\hskip 15pt}c@{\hskip 15pt}c}
\hline
\textbf{Redshift} & \textbf{Mass Median $\pm$ $\sigma$} & \textbf{SFR Median $\pm$ $\sigma$} \\
\hline
0--0.1   & $9.48 \pm 0.64$ & $2.04 \pm 0.51$ \\
0.1--0.2 & $10.14 \pm 0.39$ & $2.76 \pm 0.33$ \\
0.2--0.3 & $10.72 \pm 0.46$ & $3.37 \pm 0.29$ \\
0.3--0.4 & $10.97 \pm 0.44$ & $3.56 \pm 0.31$ \\
\hline
\end{tabular}
\end{table}

\noindent\textit{Notes.} We present the median value and the standard deviation of the stellar mass and SFR for the four redshift intervals.

\hyperref[tab1]{Table 1} quantifies the changes in stellar mass and SFR with redshift, as shown in the right panel of \hyperref[fig6]{Figure 6}. As in previous sections, the redshift is divided into four intervals to observe the corresponding changes in stellar mass and SFR. As predicted, for the data from this investigation, both stellar mass and SFR increase steadily with redshift. It should be noted that observational biases may affect the z-SFMS relationship. At higher redshifts, galaxies that can be detected tend to be brighter, while fainter galaxies are harder to observe, leading to the possible exclusion of low-mass, high-redshift galaxies. The brighter galaxies are associated with a higher number of active O-type and B-type stars, which leads to a higher SFR. Consequently, galaxies at higher redshifts tend to have higher stellar masses and higher SFRs. To mitigate the effects of these factors on the interpretation of the relationship between redshift and the nebula ratio, the Nebula Ratio is normalized by the stellar mass to balance these influences. One way to mitigate these observational biases is to match the luminosities of high-redshift galaxies with those of various mass galaxies at lower redshifts, thus achieving a controlled comparison.

\vspace{-6pt}
\begin{figure}[H]
    \centering
    \includegraphics[width=1\linewidth]{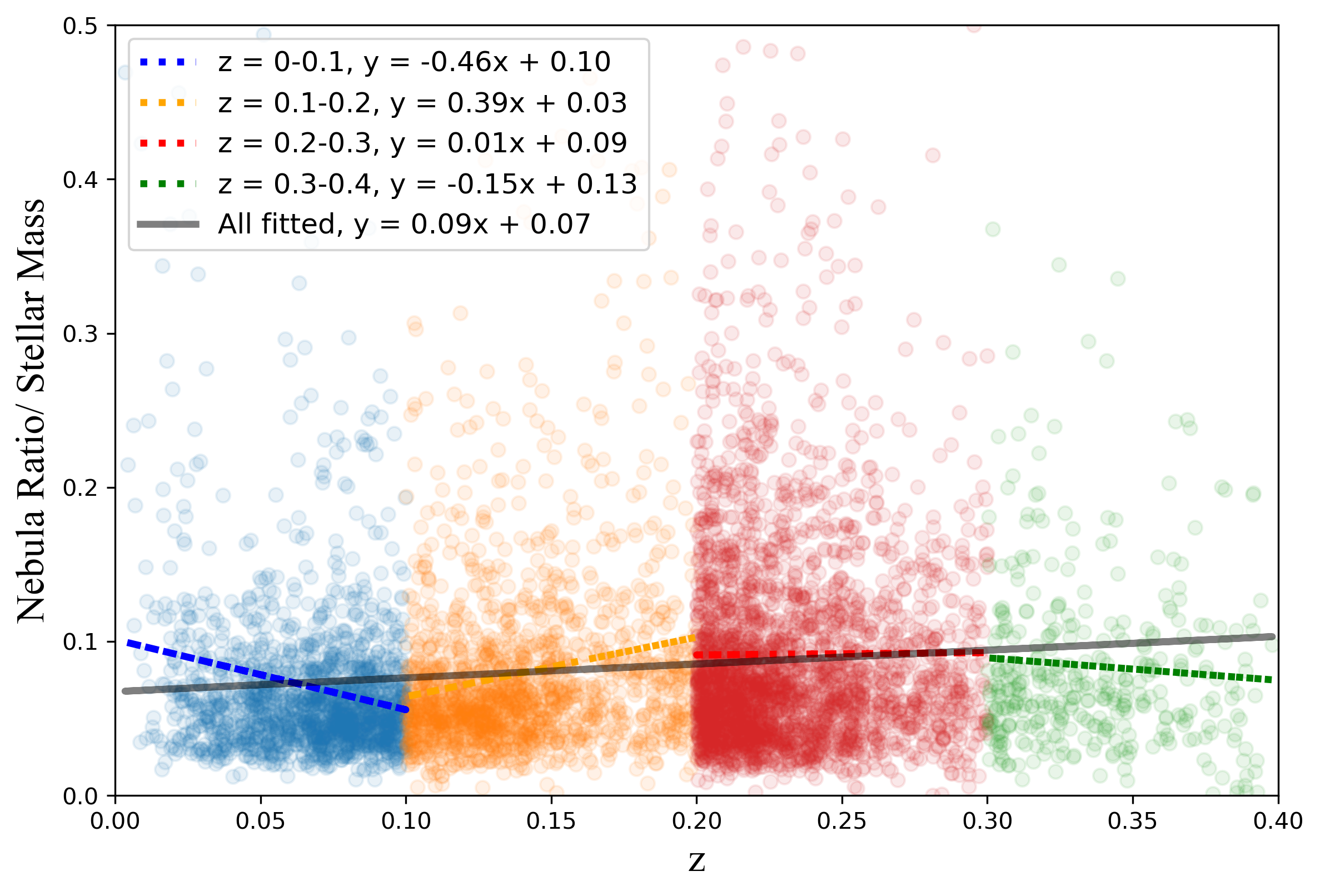}\vspace{-6pt}
    \caption{Eliminate the effects of Stellar Mass.}
    \label{fig8}
\end{figure}

\hyperref[fig8]{Figure 8} offsets the effect of stellar mass on the nebula ratio. Compared with \hyperref[fig6]{Figure 6} panel (b), the overall trend does not seem to change significantly, and the trend of the nebula ratio with increasing redshift weakens. However, the dispersion is still large, which may be related to the distribution of gas in space.

\subsection{Redshift VS Nebula Ratio}\label{Subsection3.4}

\begin{figure}[H]
    \centering
    \includegraphics[width=1\linewidth]{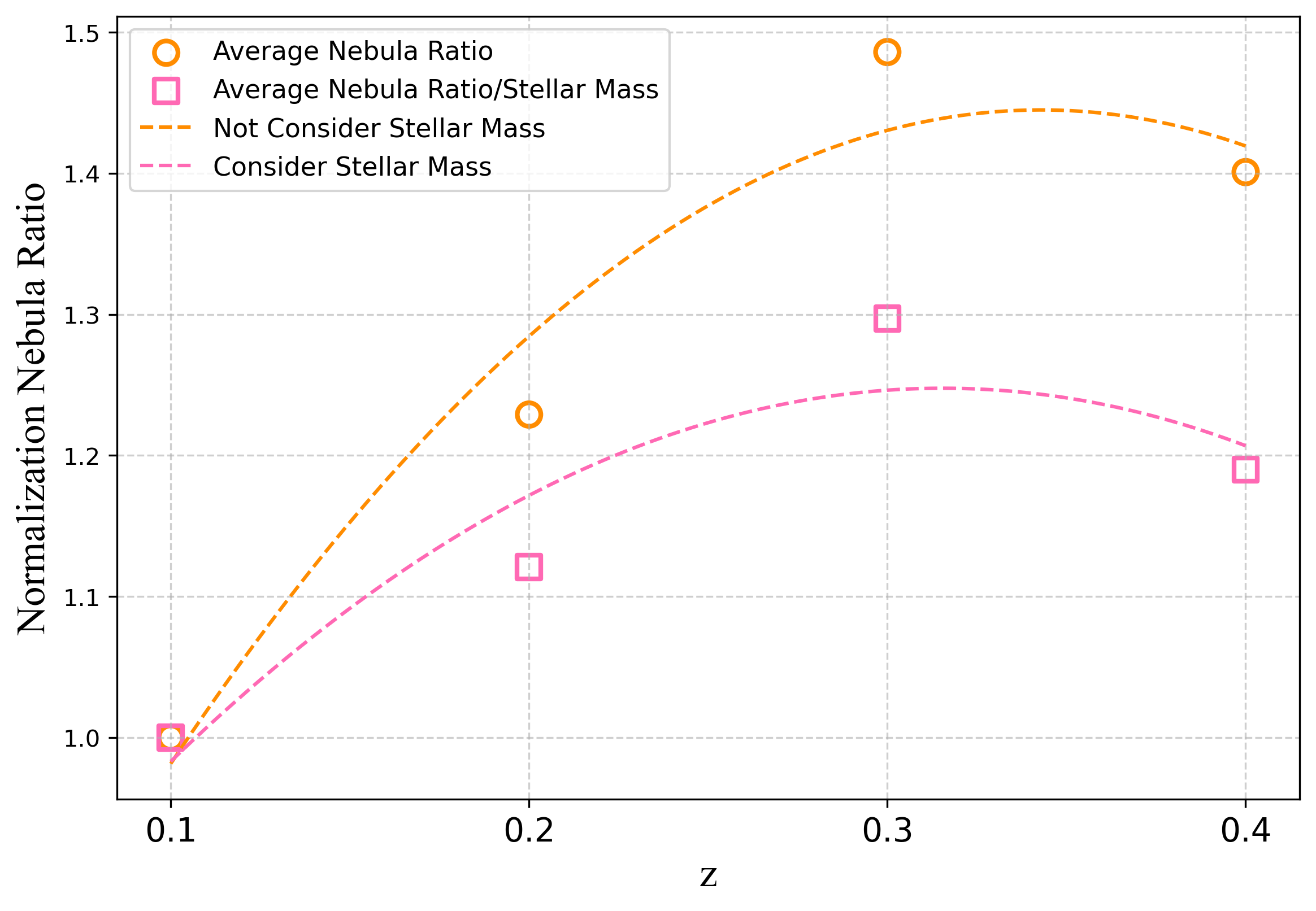}\vspace{-6pt}
    \caption{The relationship between redshift and average nebula ratio. The circular data points represent the data without considering the influence of stellar mass, and the square data points represent the data with considering the influence of stellar mass. The yellow and pink trend lines represent the quadratic equation fit of the two sets of data.}
    \label{fig9}
\end{figure}

\hyperref[fig9]{Figure 9} quantifies the impact of redshift on the average nebula ratio for the current SFG sample while also considering the influence of stellar mass. The vertical axis has been normalized. The yellow data points represent the case where the effect of stellar mass is not taken into account, while the pink data points indicate the scenario where the stellar mass effect is considered. The corresponding analytical expressions are presented in \hyperref[eq5]{Equation (5)}.

\begin{equation}
           y = ax^2 + bx + c.
\end{equation}\label{eq5}
\hyperref[eq5]{Equation (5)} displays the standard quadratic function fitting expression. The quadratic function fitting provides the best results based on the tests conducted. For the orange trend line, which does not consider the stellar mass, the coefficients are \( a = -7.853 \), \( b = 5.386 \), and \( c = 0.521 \). For the pink trend line, which accounts for the stellar mass, the coefficients are \( a = -5.700 \), \( b = 3.596 \), and \( c = 0.680 \). By comparing the \( R^2 \) values of these two datasets, it is observed that the \( R^2 \) value for the data points without the consideration of stellar mass is 0.950, whereas the \( R^2 \) value for the data points with the stellar mass effect is 0.877. Therefore, it is concluded that the impact of stellar mass on the nebula ratio is relatively small.

\vspace{-6pt}

\section{discussions} \label{section4}
This section discusses four main issues, which will serve as a foundation for future research in this area. Firstly, we compare our results with the work of \cite{Miranda2023} to explain the significant differences between our main conclusions and the relevant parts of their work. Then, we attempt to identify other factors that could affect SFR estimation. In particular, we also discuss the impact of recent significant findings on the initial mass function on SFR estimation and its trend with redshift. Finally, we look forward to future joint analyses of higher-quality spectroscopic surveys and multi-band spectroscopy.

\subsection{Comparison with Miranda's results and uncertainty} \label{subsection4.1}
Compared to the results of \cite{Miranda2023}, who used an integrated MPA-JHU sample and found negligible differences in SFR, this work leverages the latest SDSS-DR18 comprehensive galaxy data. Through a rigorous selection criterion applied via SQL queries and BPT diagrams, we ensured that all data points were purely star-forming galaxies (SFG). This guarantees that all data points exhibit significant variations in nebular emission contamination. Additionally, the redshift range in this study extends from 0 to 0.4, surpassing the range of 0 to 0.3 used by Miranda et al. (2023), where data beyond a redshift of 0.25 is sparse. Although many galaxy spectra with poor quality in the redshift range of 0.3–0.4 were excluded during the fitting process, this range represents the upper limit for accurately estimating SFR using $H_{\alpha}$ in the SDSS survey. By extending the redshift range of the galaxies to a maximum of 0.4, preliminary evidence suggests that FADO exhibits less pronounced removal of nebular emission levels at low to intermediate redshifts. Moreover, there are relatively obvious changes in the presence of nebular contamination in the SFR calculations as redshift increases. SFMS further indicates that the stellar mass and SFR obtained with FADO are closer to theoretical predictions. Due to the larger uncertainties of FADO, we observed discrepancies between the two codes in the redshift range of 0–0.1, as shown in \hyperref[fig5]{Figure 5}. However, we attribute these differences to variations in the fitting procedures of the codes rather than the effects of nebular emission contamination. The following section will analyze the reasons for the fitting uncertainties in greater detail.

Both FADO and {\it qsofitmore} employ the MC method to obtain uncertainties in emission lines, but some dominant uncertainties are still inherent in their respective codes. The large uncertainties in FADO arise from its high degrees of freedom, and the uncertainties calculation method is relatively simple, involving the superposition of components. FADO fitting includes a greater variety of spectral fitting models and considers a wider range of components. As more statistical uncertainties and degeneracies are introduced among physical components, comparing the new and old codes becomes increasingly complex \citep{Cardoso2022}. The uncertainties in FADO are also derived from random scattering in the MC model, with each step of FADO fitting generating an uncertainty caused by MC. Therefore, for FADO, a comprehensive and complex self-consistent code incorporating many spectral fitting scenarios inevitably generates more uncertainties. When these uncertainties are superimposed, significant uncertainties are observed in the figures from the previous section.  Additionally, the number of random scattering points in MC still needs to be considered. FADO iterates the Monte Carlo algorithm more times, while for {\it qsofitmore}, we only iterated 15 times. More iterations bring uncertainties closer to reality, but excessive iterations exponentially increase the time taken for {\it qsofitmore} to process spectral fitting. To avoid lengthy spectral fitting times, we chose a relatively reasonable value. 

\subsection{Other factors affecting SFR estimation} \label{subsection4.2}
The $H_{\alpha}$ obtained by FADO is generally similar to that of {\it qsofitmore}, with a median ratio of $\sim$0.0146 dex. For the SED model used by FADO, factors such as electron density, extinction, temperature, and nebular kinematics influence the nebular term in the process of fitting spectra with FADO, with electron density and temperature being the dominant factors \citep{Gomes2017}. Therefore, the underestimation of SFR derived from FADO in this study may be partly attributed to the overestimation of electron density and temperature. Accurate estimation and constraints of galaxy SFR require the combination of multi-band spectral observations due to the distribution of ultraviolet, optical, and infrared radiations in different regions of galaxies and dust obscuration \citep{Elbaz2018}. AGNs are considered to play a crucial role in shaping and evolving galaxies. The gas on the host galaxy disk interacts with AGN feedback, strongly influencing the overall galaxy SFR. AGNs that may exist in some starburst galaxies can trigger, enhance, or suppress star formation \citep{Elbaz2018}. This also affects the accuracy of estimating galaxy SFR, regardless of the spectral band used.

\subsection{The impact of variable IMF on spectral fitting} \label{subsection4.3}
Current galaxy spectral fitting (e.g., FADO and \textit{{\it qsofitmore}}) assumes the initial mass function (IMF) of stars to be constant. However, a recent discovery suggests that the IMF may not be a completely invariant cornerstone of spectral fitting theories. \citet{Li2023} recently announced that the IMF is a function primarily dependent on stellar metallicity rather than a constant as previously assumed. Since stellar metallicity varies among galaxies at different redshifts, the IMF becomes a function of redshift. Thus, future galaxy spectral fitting may need to consider the evolutionary trend of the IMF with redshift. This could profoundly affect the accurate estimation of galaxy SFRs and impose an additional redshift-evolution trend closely related to the IMF on the trend of SFR evolution with redshift.

The impact of the IMF on this survey can be described as follows: (a) The luminosity of $H_{\alpha}$ is influenced by the IMF. The Lyman continuum (LyC) photon rate is strongly dependent on redshift, and the intensity of nebular emission lines remains a topic of ongoing debate due to inaccurate redshifts. If the Salpeter IMF is used, the luminosity of $H_{\alpha}$ will vary by a factor of 4.6 from z = 0.001 to z = 0.040 \citep{Weilbacher2001}. Therefore, calculating SFR using the luminosity of $H_{\alpha}$ can lead to significant uncertainties. (b) The influence of metallicity on SFR. Low-metallicity stellar populations are brighter and hotter than solar-metallicity stellar populations with the same IMF and mass constraints. According to the survey by \citet{Weilbacher2001}, estimating the luminosity of $H_{\alpha}$ in low-metallicity dwarf galaxies using empirical calibration may result in an overestimation of the star formation rate by a factor of 3 due to the metallicity effect. (c) Metallicity affects the luminosity of $H_{\alpha}$, thereby affecting SFR: a comparison between the IMF of Scalo (1986) and Salpeter (1955) revealed a $35\%$ decrease in $H_{\alpha}$ luminosity \citep{Gomes2017}. (d) The impact of electron density (ne) on SFR. A drawback of all photoionization spectral synthesis (pss) codes is neglecting ne, which is particularly crucial for modelling pss \citep{Gomes2017}. Higher electron density implies a higher probability of gas ionization, resulting in more ionized gas. Collisions between them lead to transitions in electron energy levels, ultimately affecting the estimation of SFR.

However, IMF remains an outstanding question in astrophysics to date. The variation in IMF does not seem as straightforward as initially imagined. \citet{Andre2010}'s survey of nearby clouds indicates that the core mass function appears significantly steeper than the cloud mass function and is closer to the mass function of stars. Recently, \citet{Li2023} announced that the IMF is a primary function dependent on stellar metallicity rather than a constant function, as previously believed. Since stellar metallicity varies among galaxies at different redshifts, the IMF becomes a function of redshift. Therefore, future galaxy spectral fitting may need to consider the evolutionary trend of the IMF with redshift. The calibration by \citet{Kennicutt1998} employs model combinations from the literature and assumes a single power-law IMF with mass limits of 0.1 and 100 $M_{\text{Sun}}$ \citep{Salpeter1955}. Compared to a more realistic IMF for $H_{\alpha}$, this IMF provides a satisfactory SFR calibration, but for other wavelengths, the relative calibration using different tracers is sensitive to the precise form of the IMF. Therefore, for this survey, we still adopt the IMF form of BC03 and $H_{\alpha}$ as the SFR tracer. Different IMFs will affect estimates of star formation rates and stellar mass. However, this effect can be directly addressed by applying a conversion factor between estimates obtained with two IMFs \citep{Zahid2012}.

\subsection{Future spectroscopic data} \label{subsection4.4}
This work ultimately obtained the expected results; preliminary evidence suggests that estimating the SFR for higher redshift galaxies using $H_{\alpha}$ is minimally affected by nebular emission contamination. However, this investigation also has a limited sample size for SFR calculations. In addition to expecting improvements in the quality of future SDSS spectral data, supplementary studies using spectral surveys such as DESI have potential. Currently, the S/N of the spectra of the majority of SDSS galaxies with redshifts higher than 0.4 is very low. The data quality is poor, making it difficult to identify and fit emission lines related to SFR estimation. Therefore, if exploring the SFR of galaxies at higher redshifts, especially those in the cosmic noon period at z $\sim$ 2, not only optical spectra but also UV or IR spectra need to be combined, which will require a huge amount of work. However, there are still doubts about whether FADO has good IR spectral analysis capabilities. We hope that the FADO team can improve and test the code of FADO in the future to perform well not only in IR but also in UV spectra. With the extensive implementation of galaxy IR observation projects by JWST, the era of large sample analysis of high-redshift galaxies will soon arrive. The question of the SFR of the first-generation galaxies in the universe will also be revealed.

In addition, it is imperative to mention that estimating the SFR of a galaxy through multi-band observations leads to more precise results. The estimation of SFR via $H_{\alpha}$ emission is also constrained; $H_{\alpha}$ is mostly applicable to galaxies with mean ages ranging from 0.3 to 10 million years (Myr). Comparatively, utilizing 2-10 keV may yield more effective results than $H_{\alpha}$. Therefore, for accurate SFR computation, employing multi-band observations is essential to obtain the most comprehensive spectrum of the target galaxy, thereby deriving a complete SFR. Further weight is applied to alternative estimations of attenuation correction and SFR based on UV and IR measurements \citep{Kennicutt2012}. Additionally, disregarding the impact of the IMF, significant differences are evident in SFR calculations at $H_{\alpha}$ luminosities of $10^{38}$ \citep{Cervino2003}. Neglecting the IMF effect for lower SFRs may result in the misestimation of stellar lifetimes, thereby elongating their lifecycle and inducing larger fluctuations in $H_{\alpha}$ emission lines. Hence, it is advisable to employ tracers other than the emission line $H_{\alpha}$ to constrain SFR, such as FUV emission \citep{Kennicutt2012}. Observations in the infrared band are crucial for obtaining SFR, as interstellar dust absorbs half of the starlight in galaxies and re-emits it in the infrared. The morphological variations in different dust emission components translate into significant changes in the spectral energy distribution (SED) of dust within and between galaxies \citep{Smith2007}; thus, the conversion of SFR from multi-band observations still requires calibration based on different bands. Regarding correcting SFR after combining optical and infrared observations, \citet{Rosario2016, Kennicutt2012} have made notable contributions.

Therefore, to obtain the most accurate SFR for galaxies, combining the target galaxy's visible, infrared, and ultraviolet spectra into a complete galaxy spectrum is necessary. However, the objective of this investigation is to compare the effects of two different fitting methods on SFR. Hence, the aforementioned impacts have fairly appeared in both fitting methods, cancelling each other out. Nevertheless, it is crucial to note that obtaining accurate SFR requires the combination of correct IMF and multi-band observations.

\section{Conclusion} \label{section5}
In this preliminary study, we used the latest large-scale SDSS-DR18 survey data, applying stringent criteria to restrict the entire sample to SFG. The redshift range is restricted by the limit for estimating SFR using $H_{\alpha}$. Beyond a redshift of 0.4, the quality of $H_{\alpha}$ emission is poor, and the estimated SFR becomes less accurate. This survey further validates the findings of \cite{Miranda2023}, which show that the contribution of nebular emission to the star formation rate increases with redshift. This insight could be valuable for future studies on star formation rates at high redshifts, especially for starburst galaxies and SFGs, where the contribution of nebulae cannot be overlooked. The three main conclusions of this survey are presented below:

(1) The pure stellar code {\it qsofitmore} and the FADO code, which accounts for the nebular emission, show consistent results in comparison of the $H_{\alpha}$ flux. The middle difference is \( 0.034 \, \text{dex} \approx 8.16\% \). There is no significant trend in the differences between the two codes concerning the redshift. Therefore, for SFGs with \( z < 0.4 \), it seems unnecessary to consider nebular emission in the spectral fitting process. However, for individual source analyses, considering nebular emission results in a reduced $H_{\alpha}$ flux.

(2) For the trend of nebula ratio with redshift, there is an overall upward trend, particularly in the range of \( z = 0.1 - 0.3 \). However, the data points are dispersion, and no clear trend can be concluded. However, the nebula ratio indeed increases with redshift, and more data will be needed to confirm this observation.

(3) The average nebula ratio for the four redshift bins shows a relatively stable upward trend with redshift, and \hyperref[eq5]{Equation (5)} is used to quantify this relationship. We also found that the nebula ratio does not seem to have a strong correlation with stellar mass. For SFGs with \( z < 0.4 \), the impact of stellar mass can be somewhat neglected.

Finally, we predict that the difference in nebular emission between FADO and other purely stellar synthesis codes will become even more pronounced at higher redshifts. However, it remains essential to investigate the effect of FADO on the fitting of high-redshift galaxies and the impact of using other emission lines on SFR estimation.

\section*{Acknowledgement}
The author thanks the anonymous referee for the detailed reading of the paper and useful constructive suggestions and comments. This work has used Sloan Digital Sky Survey VIII spectroscopic data products. The Alfred P. Sloan Foundation, the U.S. Department of Energy Office of Science, and the Participating Institutions have funded the Sloan Digital Sky Survey VIII. SDSS-VIII acknowledges support and resources from the Center for High-Performance Computing at the University of Utah. The SDSS website is www.sdss.org, and the spectroscopic data website used in this work is \url{https://dr18.sdss.org/optical/spectrum/search}. 

The author sincerely thanks Jiahe Xiao for helping with the FADO running problem. The author thanks Han Yirui for all your support. 

\bibliographystyle{aasjournal}

\begin{appendix}
\section{List of SQL query}
\label{AppendixA}
To find suitable SFG with an obvious velocity offset (velocity offset $> 150 \, \mathrm{km \, s^{-1}}$), we select galaxies in SDSS DR18 \citet{Almeida2023} with the Structured Query Language (SQL) Search tool (\url{http://skyserver.sdss.org/dr18/en/tools/search/sql.aspx}). The applied SQL query in detail is as follows:
\noindent
\begin{minipage}{0.48\textwidth}
\textbf{SQL Query 1}:
\begin{verbatim}
SELECT S.plate, S.fiberid, S.mjd, 
S.z,G.oiii_5007_flux,
G.h_beta_flux, G.h_beta_flux_err,
G.oiii_5007_flux, G.oiii_5007_flux_err,
G.nii_6584_flux, G.nii_6584_flux_err,
G.h_alpha_flux, G.h_alpha_flux_err,
G.oiii_5007_flux/G.h_beta_flux as o3hb,
G.nii_6584_flux/G.h_alpha_flux as n2ha
FROM GalSpecLine as G
JOIN SpecObjall as S
ON S.specobjid = G.specobjid
WHERE S.class ='GALAXY' and
S.SNmedian >5 and SNmedian < 20
and S.z <0.2 and z >=0 and S.zwarning = 0 and
S.veldisperr > 0
and S.veldisp > 3*S.veldisperr and
(S.veldisp > 80 and S.veldisp <350)
and G.h_beta_flux_err > 0 and
G.h_beta_flux >5*G.h_beta_flux_err
and G.h_alpha_flux_err > 0 and
G.h_alpha_flux > 5*G.h_alpha_flux_err
and G.oiii_5007_flux_err > 0 and
G.oiii_5007_flux >5*G.oiii_5007_flux_err
and G.nii_6584_flux_err > 0 and
G.nii_6584_flux > 5*G.nii_6584_flux_err
\end{verbatim}
\end{minipage}
\hfill
\begin{minipage}{0.48\textwidth}
\textbf{SQL Query 2}:
\begin{verbatim}
SELECT S.plate, S.fiberid, S.mjd, 
S.z, G.oiii_5007_flux,
G.h_beta_flux, G.h_beta_flux_err,
G.oiii_5007_flux, G.oiii_5007_flux_err,
G.nii_6584_flux, G.nii_6584_flux_err,
G.h_alpha_flux, G.h_alpha_flux_err,
G.oiii_5007_flux/G.h_beta_flux as o3hb,
G.nii_6584_flux/G.h_alpha_flux as n2ha
FROM GalSpecLine as G
JOIN SpecObjall as S
ON S.specobjid = G.specobjid
WHERE S.class ='GALAXY' and
S.SNmedian >5
and S.z <0.4 and z >=0.2 and S.zwarning = 0 and
S.veldisperr > 0
and S.veldisp > 3*S.veldisperr and
(S.veldisp > 80 and S.veldisp <350)
and G.h_beta_flux_err > 0 and
G.h_beta_flux >5*G.h_beta_flux_err
and G.h_alpha_flux_err > 0 and
G.h_alpha_flux > 5*G.h_alpha_flux_err
and G.oiii_5007_flux_err > 0 and
G.oiii_5007_flux >5*G.oiii_5007_flux_err
and G.nii_6584_flux_err > 0 and
G.nii_6584_flux > 5*G.nii_6584_flux_err
\end{verbatim}
\end{minipage}

\section{Example of {\it qsofitmore} \& FADO spectral fitting}
\label{AppendixB}

This section shows the comparison between {\it qsofitmore} and FADO fitting results for the same galaxy spectrum LEDA 1171303.
\vspace{-8pt}

\begin{figure}[htbp]
    \centering
    \includegraphics[width=0.8\linewidth]{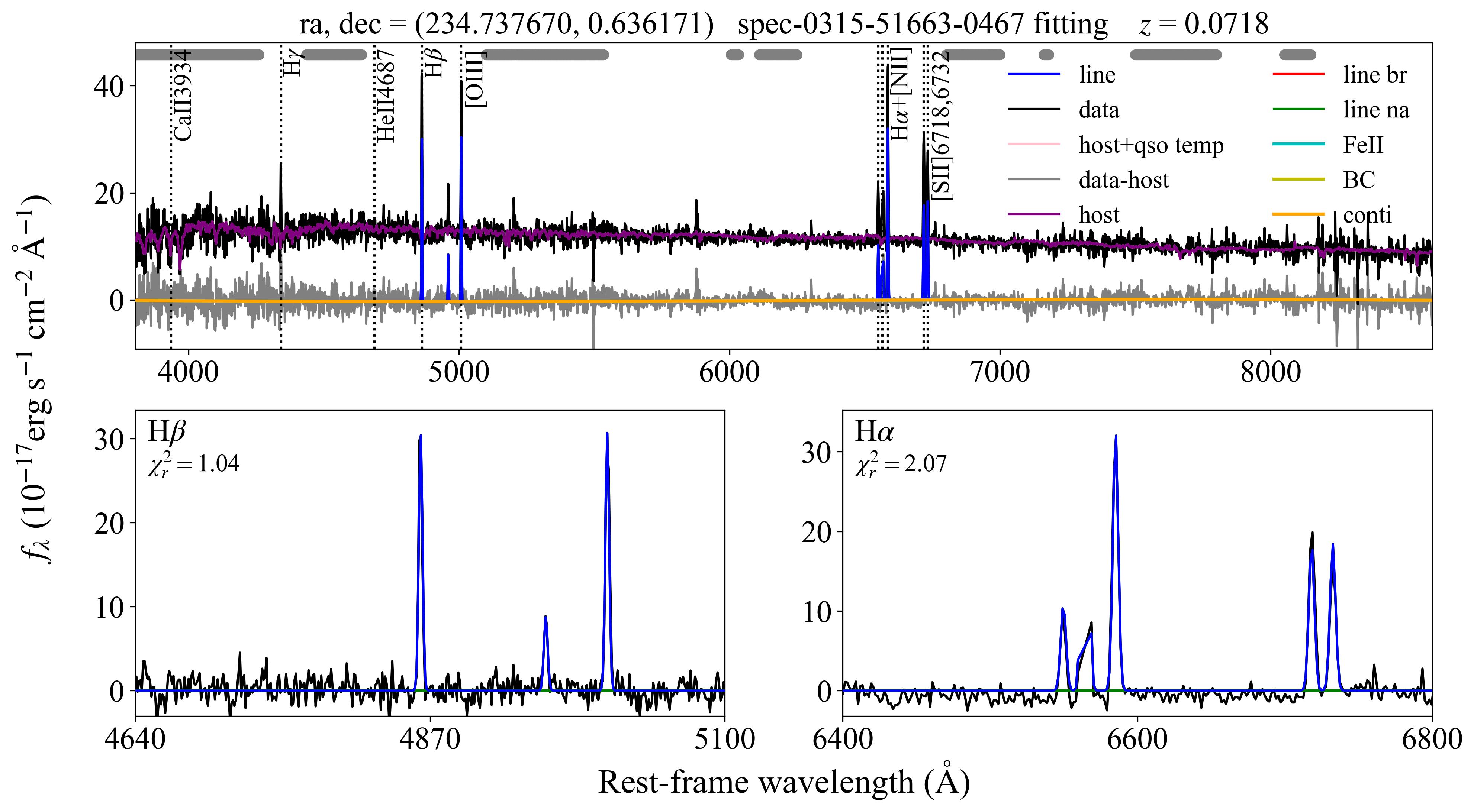}
    \vspace{-6pt}
    \caption{
        The fitting result of galaxy LEDA 1171303 with {\it qsofitmore}. 
        \textit{Upper panel}: The fitting of the full spectrum considering the subtraction of the host component. 
        \textit{Lower panels}: Zoom-in on the fitting of $H_{\alpha}$ and $H_{\beta}$.
    }
    \label{fi10}
\end{figure}

\begin{figure}[htbp]
    \centering
    \rotatebox{0}{
        \includegraphics[width=0.7\linewidth]{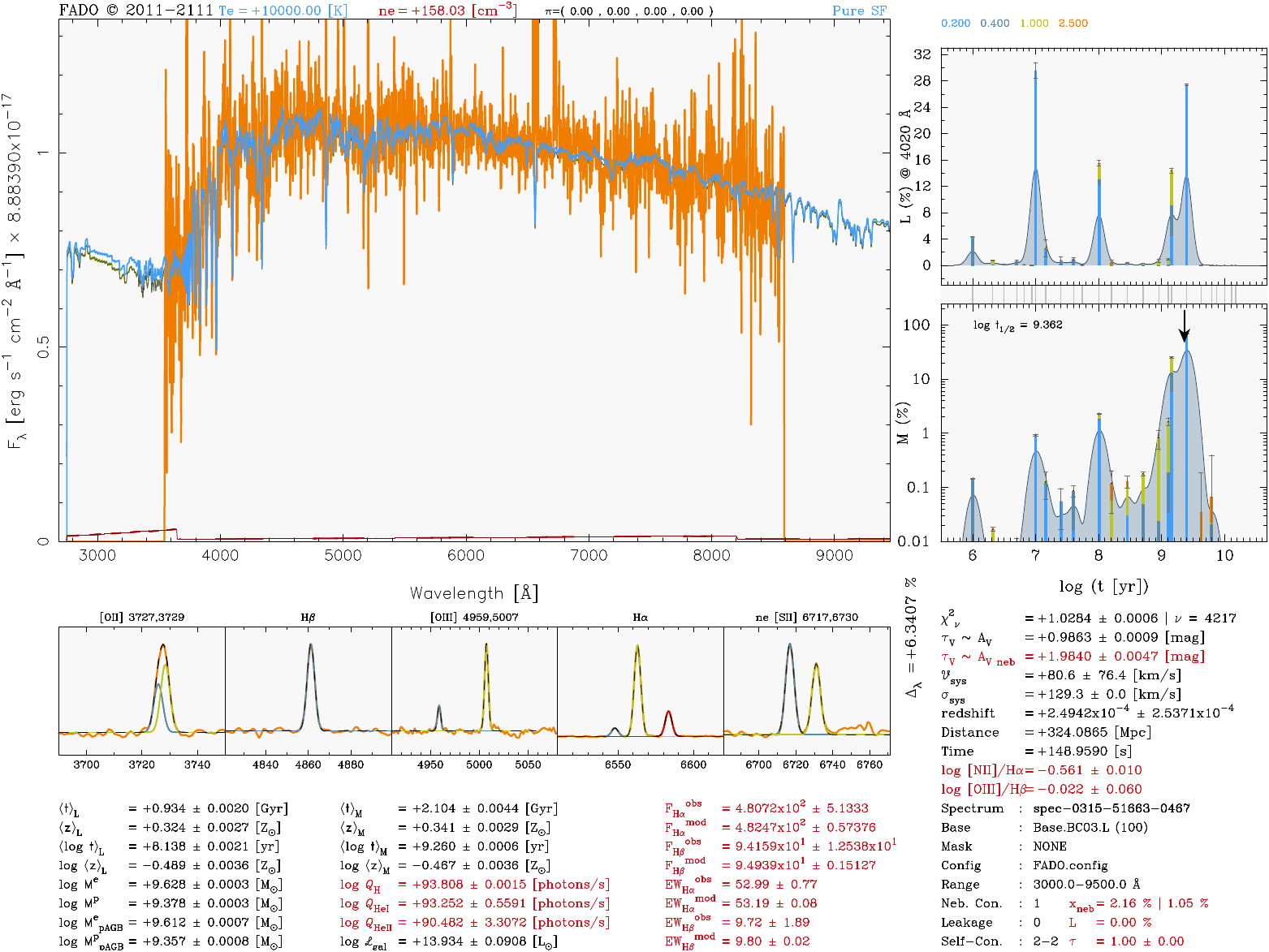}
    }
    \vspace{-6pt}
    \caption{
        The fitting result of galaxy LEDA 1171303 with FADO. 
        \textit{Lower panels}: The main emission line fitted result, the red line at the bottom representing the nebula model flux. The yellow line represents the input data, and the green line is the stellar model flux. The blue line is the total fitting model.
        For a more detailed analysis of this fit, see \citet{Gomes2017}.
    }
    \label{fig11}
\end{figure}

\end{appendix}

\end{document}